\documentstyle[preprint,aps,psfig]{revtex}

\def\beq{\begin{equation}}
\def\eeq{\end{equation}}
\def\beqn{\begin{eqnarray}}
\def\eeqn{\end{eqnarray}}

\begin{document} 

\draft 
\preprint{MKPH-T-98-19}

\title {The Generalized Gerasimov-Drell-Hearn Integral\\
 and the Spin Structure of the Nucleon} 

\author{D. Drechsel, S.S. Kamalov\thanks{ Permanent address: 
Laboratory of Theoretical Physics, JINR Dubna, 
Head Post Office Box 79, SU-101000 Moscow, Russia.}, 
G. Krein, and L. Tiator}

\address{Institut f\"ur Kernphysik, Johannes Gutenberg-Universit\"at, D-55099
Mainz, Germany}

\maketitle 

\begin{abstract}
The spin structure functions $g_1$ and $g_2$ have been calculated
in the resonance region and for small and intermediate momentum
transfer. The calculation is based on a gauge-invariant and 
unitary model for one-pion photo- and electroproduction. The
predictions of the model agree with the asymmetries and the
spin sturcture functions recently measured at SLAC, and the
first moments of the calculated spin structure functions fullfil
the Gerasimov-Drell-Hearn and Burkhardt-Cottingham sum rules
within an error of typically 5-10~$\%$.
\end{abstract}


\section{Introduction}
The spin structure of the nucleon has been studied by increasingly
accurate measurements since the end of the 70's. By scattering
polarized lepton beams off polarized targets, it has become 
possible to determine the spin structure functions $g_1(x,Q^2)$
and $g_2(x,Q^2)$. These functions depend on the fractional
momentum of the proton, $x=Q^2/2m\nu$, where $-Q^2$ is the square
of four-momentum transfer, $\nu$ the energy transfer in the lab
frame, and $m$ the nucleon mass. The results of the first 
experiments at CERN~\cite{Ash89} and SLAC~\cite{Bau83} sparked
considerable interest in the community, because the first moment
of $g_1$, $\Gamma_1=\int^1_0g_1(x)dx$, was found to be 
substantially smaller than expected from the quark model, in
particular from the Ellis-Jaffe sum rule~\cite{Ell74}. This 
``spin crisis'' of the nucleon led to the conclusion that only
a small fraction of the nucleon's spin is carried by the
valence quarks, and that the main spin contributions should be
due to sea quarks, quark orbital momentum and gluons. However,
the difference of the proton and neutron moments,
$\Gamma_1^p-\Gamma_1^n$, was found to be well described by
Bjorken's sum rule~\cite{Bjo66}, which is a strict QCD prediction.

Recent improvements in polarized beam and target techniques have
made it possible to determine both spin structure functions over
an increased range of kinematical values. The reader is referred
to the publications of the collaborations at CERN~\cite{Ada97},
SLAC~\cite{Abe97} and DESY~\cite{Ack97} and to the references
on earlier work quoted therein. While most of these investigations
have concentrated on the high $Q^2$ region, the recent measurements
of the E143 Collaboration at SLAC~\cite{Abe98} have been
extended to the range of
0.6 (GeV/c)$^2\le Q^2\le$ 1.2 (GeV/c)$^2$, which bridges the
gap between low-$Q^2$ nonperturbative resonance phenomena and
high-$Q^2$ perturbative QCD.

It is the aim of our contribution to present the results of the
recently developed Unitary Isobar Model (UIM, Ref.~\cite{Dre98a}) for
the spin asymmetries, structure functions and relevant sum rules in
the resonance region. This model describes the presently available
data for single-pion photo- and electroproduction up to a total $cm$
energy $W_{\rm max}\approx 2$ GeV and for $Q^2\le$ 2 (GeV/c)$^2$. It
is based on effective Lagrangians for Born terms (background) and
resonance contributions, and the respective multipoles are constructed
in a gauge-invariant and unitary way for each partial wave.  The eta
production is included in a similar way~\cite{Kno95}, while the
contribution of more-pion and higher channels is modeled by comparison
with the total cross sections and simple phenomenological assumptions.

The spin structure in the resonance region has intrigued many authors
because of an expected rapid change from the physics of
resonance-driven coherent processes to incoherent scattering off the
constituents. In particular the first moment $\Gamma_1$ is
constrained, in the limit of $Q^2\rightarrow 0$ (real photons), by the
famous Gerasimov-Drell-Hearn sum rule (GDH, Ref.~\cite{Ger65}),
$\Gamma_1\rightarrow -Q^2\kappa^2/8m^2$, where $\kappa$ is the
anomalous magnetic moment of the nucleon.  The reader should note that
here and in the following we have included the inelastic contribution
to $\Gamma_1$ only. As has been pointed out by Ji and
Melnitchouk~\cite{JiM97}, the elastic contribution is in fact the
dominant one at small $Q^2$ and has to be taken into account in
comparing with twist expansions about the deep-inelastic limit at
those values of $Q^2$.

In the case of the proton, the GDH sum rule predicts $\Gamma_1<0$ for
small $Q^2$, while all experiments for $Q^2>1$ (GeV/c)$^2$ yield
positive values. Clearly, the value of the sum rule has to change
rapidly at low $Q^2$, with some zero-crossing at $Q^2_0<1$
(GeV/c)$^2$. The evolution of the sum rule was first described by
Anselmino et al.~\cite{Ans89} in terms of a parametrization based on
vector meson dominance. Burkert, Ioffe and others~\cite{Bur92,Iof97} refined
this model considerably by treating the contributions of the
resonances explicitly. Soffer and Teryaev~\cite{Sof93} suggested that
the rapid fluctuation of $\Gamma_1$ should be analyzed in conjunction
with $\Gamma_2$, the first moment of the second (transverse) spin
function. The latter is constrained by the less familiar
Burkhardt-Cottingham sum rule (BC, Ref.~\cite{Bur70}) at all values of
$Q^2$. Therefore the sum of the two moments, $\Gamma_1 +\Gamma_2$, is
known for both $Q^2=0$ and $Q^2 \rightarrow \infty$. Though this sum
is related to the practically unknown longitudinal-transverse
interference cross section $\sigma'_{LT}$ and therefore not yet
determined directly, it can be extrapolated smoothly between the two
limiting values of $Q^2$. The rapid fluctuation of $\Gamma_1$ then
follows by subtraction of the BC value of $\Gamma_2$. The importance
of $\sigma'_{LT}$ for a complete understanding of the spin structure
has also been stressed by Li and Li~\cite{LiL94}.

The GDH sum rule is obtained by integrating the quantity
$(\sigma_{1/2}-\sigma_{3/2})/\nu$ over the photon energy, where
$\sigma_{3/2}$ and $\sigma_{1/2}$ are the cross sections for virtual
photon and target spins in parallel and antiparallel, respectively.
Several authors have pointed out that serious discrepancies exist
between the sum rule and theoretical analysis based on the existing
unpolarized data. The small momentum evolution of the extended GDH sum
rule was also investigated in the framework of heavy baryon
ChPT~\cite{Ber93}. The authors predicted a positive slope of the GDH
integral for $Q^2=0$, while the phenomenological analysis of Burkert
et al.~\cite{Bur92} indicated a negative slope. However, the
integrands of all spin-related sum rules are fluctuating functions of
energy with both positive and negative contributions. Therefore, even
seemingly small uncertainties may give rise to large differences (see
Ref.~\cite{Dre98b} and references quoted therein).

The first direct experimental data have been recently taken at
MAMI~\cite{Ahr92} in the energy range 200 MeV$<\nu<$ 800 MeV, and data
at the higher energies are expected from ELSA within short. Concerning
electroproduction in the resonance region, the $\Delta$(1232) is being
investigated at ELSA, Jefferson Lab, MAMI and MIT/Bates ~\cite{Got98}.
The high-duty factor of these modern electron accelerators makes it
possible to measure not only the (total) helicity cross sections but
also the individual decay channels, thus providing a much more
stringent test of model descriptions.  The spin structure of the
nucleon will be studied at the Jefferson Lab by a series of
experiments~\cite{Bur91}, which are expected to cover the entire
resonance region and momentum transfers up to $Q^2\approx 2$
(GeV/c)$^2$. The outcome of these experiments should answer the open
questions about the helicity structure at low and intermediate
energies, the validity of the GDH sum rule, the size and role of the
transverse-longitudinal contribution in connection with the BC sum
rule and the generalized GDH integral, the evolution of these
integrals with momentum transfer, and the contribution of individual
decay channels to the spin structure functions.

In the following we review the formalism of spin structure functions
and sum rules in sect. 2. We then present the predictions of the UIM
in sect. 3, and close by a brief summary of our findings in sect. 4.

\section{Formalism}
 
We consider the scattering of polarized electrons off polarized target
nucleons. The $lab$ energies of the electrons in the initial and final
states are denoted with $E$ and $E'$, respectively. The incoming
electrons carry the (longitudinal) polarization $h=\pm 1$, and the two
relevant polarization components of the target are $P_z$ (parallel to
the $lab$ momentum $\vec{k}$ of the virtual photon) and $P_x$
(perpendicular to $\vec{k}$ in the scattering plane of the electron
and in the half-plane of the outgoing electron). The differential
cross section for exclusive electroproduction can then be expressed in
terms of four ``virtual photoabsorption cross sections''
$\sigma_i(\nu, Q^2)$ by~\cite{Dre94}

\beq
\frac{d\sigma}{d\Omega\ dE'} = \Gamma \sigma (\nu,Q^2)\,,
\label{eq1}
\eeq

\beq
\sigma = \sigma_T+\epsilon\sigma_L+hP_x\sqrt{2\epsilon(1-\epsilon)}\ 
         \sigma'_{LT}+hP_z\sqrt{1-\epsilon^2}\sigma'_{TT}\, ,
\label{eq2}
\eeq
with

\beq
\Gamma = \frac{\alpha}{2\pi^2}\ \frac{E'}{E}\ \frac{K}{Q^2}\ 
         \frac{1}{1-\epsilon}
\label{eq3}
\eeq 
the flux of the virtual photon field and $\epsilon$ its
transverse polarization, $\nu=E-E'$ the virtual photon energy in the
$lab$ frame and $Q^2>0$ describing the square of the virtual photon
four-momentum. In accordance with our previous notation~\cite{Dre98a}
we shall define the flux with the ``photon equivalent energy''
$K=k_{\gamma}=(W^2-m^2)/2m$, where $W$ is the total $cm$ energiy and
$m$ the mass of the target nucleon.

The quantities $\sigma_T$ and $\sigma'_{TT}$ can be expressed in terms
of the total cross sections $\sigma_{3/2}$ and $\sigma_{1/2}$,
corresponding to excitation of hadronic states with spin projections
$3/2$ and $1/2$, respectively,

\beq
\sigma_T = \frac{1}{2}(\sigma_{3/2} + \sigma_{1/2})\,,\qquad
\sigma'_{TT} = \frac{1}{2} (\sigma_{3/2}-\sigma_{1/2})\ .
\label{eq4}
\eeq

The virtual photoabsorption cross sections in Eq.~(\ref{eq2})
are related to the quark structure functions,

\begin{eqnarray}
\label{eq5}
\sigma_T& =& \frac{4\pi^2\alpha}{mK}\ F_1\ ,\nonumber \\ 
\sigma_L& =& \frac{4\pi^2\alpha}{K}\left[ \frac{F_2}{\nu}
(1+\gamma^2) - \frac{F_1}{m}\right]\ ,\nonumber \\
\sigma'_{LT}& =& -\frac{4\pi^2\alpha}{mK}\,\gamma\,(g_1+g_2)\ ,\nonumber \\
\sigma'_{TT}& =& -\frac{4\pi^2\alpha}{mK}\left (g_1-\gamma^2\, g_2\right )\ ,
\end{eqnarray}
with $\gamma=Q/\nu$, and the quark structure functions $F_1$, $F_2$,
$g_1$, and $g_2$ depending on $\nu$ and $Q^2$. In comparing with the
standard nomenclature of DIS~\cite{Abe98} we note that
$\sigma'_{LT}=-\sigma_{LT}(DIS)$ and $\sigma'_{TT}=-\sigma_{TT}(DIS)$.

We generalize the GDH sum rule by introducing the $Q^2$-dependent integral
\beq
I_1(Q^2) = \frac{2m^2}{Q^2}\int_{0}^{x_0}g_1(x,Q^2)\ dx
\rightarrow 
\left \{ \begin{array}{lll}
 - \frac{1}{4}\kappa_N^2 &{\rm for} &Q^2\rightarrow 0\\
\frac{2m^2}{Q^2}\Gamma_1 + {\cal O}(Q^{-4}) &{\rm for} &Q^2\rightarrow \infty
\end{array} \right. \,,
\label{eq6}
\eeq 
where $x=Q^2/2m\nu$ is the Bjorken scaling variable and $x_0=
Q^2/(2mm_{\pi}+m_{\pi}^2+Q^2)$ refers to the inelastic threshold of
one-pion production. In the scaling regime the structure functions
should depend on $x$ only, and $\Gamma_1=\int g_1(x) dx = const$. We
note that several other generalizations of the GDH integral have been
proposed by introducing admixtures of the second spin structure
functions $g_2$ and/or factors depending on (different) definitions of
the photon flux. While all generalizations have the same limits as in
Eq.~(\ref{eq6}), they may differ substantially at intermediate
$Q^2$~\cite{Pan97}.

For the second spin structure function the Burkhardt-Cottingham (BC) 
sum rule asserts that the integral over $g_2$ vanishes if integrated
over both elastic and inelastic contributions~\cite{Bur70}.
As a consequence one finds

\beq
I_2(Q^2) = \frac{2m^2}{Q^2}\int_{0}^{x_0}g_2(x,Q^2)\ dx
=\frac{1}{4}\frac{G_M(Q^2)-G_E(Q^2)}{1+Q^2/4m^2}\,G_M(Q^2)\,,
\label{eq7}
\eeq
i.e. the inelastic contribution for $0 < x <x_0 $ equals the 
negative value of the elastic contribution given by the $rhs$
of Eq.~(\ref{eq7}), which is parametrized  by the magnetic
and electric Sachs form factors $G_M$ and $G_E$, respectively.
These form factors are  normalized by the nucleon's magnetic
moment $\mu_N=\kappa_N+e_N$ and its charge $e_N$ in
the limit of $Q^2\rightarrow 0$, $G_M(0)=\mu_N$ and $G_E(0)=e_N$.
The BC sum rule has the limiting cases

\beq
I_2(Q^2)\rightarrow \left \{ \begin{array}{lll}
              \frac{1}{4}\mu_N \kappa_N &{\rm for} &Q^2\rightarrow 0\\
              {\cal O}(Q^{-10})         &{\rm for} &Q^2\rightarrow \infty
         \end{array} \right.\,.
\label{eq8}
\eeq

We have calculated  the integrals $I_1$ and $I_2$ in terms of the  
virtual photon cross sections,

\beq
I_1(Q^2) = \frac{m^2}{8\pi^2\alpha}\int_{\nu_0}^{\infty}
           \frac{1-x} {1+\gamma^2}
           \left (\sigma_{1/2}-\sigma_{3/2}
           -2\gamma\,\sigma'_{LT}\right )\ \frac{d\nu}{\nu}\ ,
\label{eq9}
\eeq

\beq
I_2(Q^2) = \frac{m^2}{8\pi^2\alpha}\int_{\nu_0}^{\infty}
           \frac{1-x} {1+\gamma^2}
           \left (\sigma_{3/2}-\sigma_{1/2}
           -\frac{2}{\gamma}\,\sigma'_{LT}\right )\ \frac{d\nu}{\nu}\ ,
\label{eq10}
\eeq
where $ \nu_0 =m_{\pi} + (m_{\pi}^2+Q^2)/2m$ is the threshold
$lab$ energy of one-pion production.

Since $\gamma \sigma'_{LT}={\cal O}(Q^2)$, the longitudinal-transverse
term does not contribute to the integral $I_1$ in the real photon
limit.  However, the ratio $\sigma'_{LT}/\gamma$ remains constant in
that limit and hence contributes to $I_2$. As a result we find

\beq
I_2(0) = \frac{1}{4}\kappa_N^2+\frac{1}{4}e_N\kappa_N\ ,
\label{eq11}
\eeq
with the two terms on the $rhs$ corresponding to the contributions of
$\sigma_{3/2}-\sigma_{1/2}$ and $\sigma'_{LT}$, respectively.
Equation~(\ref{eq11}) provides an interesting model-independent
constraint for the two spin-dependent virtual photon cross sections in
the real photon limit.

Finally, in order to relate our calculations with the experimental
data, we express the virtual photon asymmetries $A_1$ and $A_2$
defined in Ref.~\cite{Abe98} by the virtual photon cross sections,

\begin{eqnarray}
\label{eq12}
A_1=\frac{\sigma_{1/2}-\sigma_{3/2}}{\sigma_{1/2}+\sigma_{3/2}}
 = -\frac{\sigma'_{TT}}{\sigma_T}\,,\qquad 
A_2 = -\frac{\sigma'_{LT}}{\sigma_T}\,.
\end{eqnarray}
From the measured values for the asymmetries $A_1$ and $A_2$ and from
the total transverse cross section $\sigma_T$ we then construct the
quark structure functions $g_1$ and $g_2$ by use of Eqs.~(\ref{eq4}),
(\ref{eq5}) and (\ref{eq12}),

\begin{eqnarray}
\label{eq13}
g_1=\frac{F_1}{1+\gamma^2}\,(A_1+\gamma\,A_2)\,,\qquad
g_2=\frac{F_1}{1+\gamma^2}\,\left(-A_1+ \frac{1}{\gamma}\,A_2\right)\,.
\end{eqnarray}
In practice it is difficult to measure the two (unpolarized) cross
sections $\sigma_T$ and $\sigma_L$ separately by means of a Rosenbluth
plot. The structure function $F_1$ is then extracted from the total
(unpolarized) cross section, $\sigma_{tot}=\sigma_T+\epsilon
\sigma_L$, by use of an ``educated guess'' for the ratio
$R=\sigma_L/\sigma_T$.  Moreover, if only the longitudinal target
polarization can be measured, only the combination $A_1+\eta A_2$ can
be determined, where $\eta=\epsilon Q/(E-\epsilon E')$.  In this case
the structure function $g_1$ has to be extracted by use of additional
assumptions about the second spin structure function $g_2$.

\section{Results and Discussions}

\subsection{Total cross sections}

The recently developed Unitary Isobar Model (UIM)~\cite{Dre98a}
describes the single-pion electroproduction channel quite adequately.
However, for energies $W>1.3\,GeV$ the contributions from other
channels, in particular from multipion production, become increasingly
important. At present several theoretical models are available for the
two-pion photoproduction channels~\cite{Gom94,Mur95,Och97}, but little
information exists on multipion electroproduction. In this paper we
extract the necessary information from the available data for the
total cross sections of real and virtual photon absorption on the
proton~\cite{Mac96,Bra76}, assuming that the single-pion channel is
well determined by the UIM.  In addition we account for the well
investigated eta production channel by using the parametrization for
the corresponding total cross section $\sigma_T^{(\eta)}(W,0)$ at the
photon point as suggested in Ref.\cite{Kno95}.  We extend this
parametrization for the case of virtual photons using the helicity
amplitude $A_{1/2}^{(S_{11})}$ for the $S_{11}(1535)$ resonance
obtained within the UIM, i.e.

\begin{eqnarray}
\label{eq14}
  \sigma_T^{(\eta)}(W,Q^2)=\sigma_T^{(\eta)}(W,0)\,
\left[ A_{1/2}^{(S_{11})}(Q^2)/A_{1/2}^{(S_{11})}(0)\right]^2\ .
\end{eqnarray}

The transverse part of the total cross section for multipion 
production is now obtained by

\begin{eqnarray}
\label{eq15}
   \sigma^{(n\pi)}_T(W,Q^2)=F^2(Q^2)\,\left[\sigma^{(exp)}_T(W,0) 
  - \sigma^{(1\pi)}_T(W,0) - \sigma^{(\eta)}_T(W,0)\right]
\end{eqnarray}
with a monopole form factor $F(Q^2)=1/(1+\alpha_{n\pi}\,Q^2)$
describing the $Q^2$ dependence. A value of
$\alpha_{n\pi}=1.05\,(c/GeV)^2$ provides the best fit to the total
inelastic cross section up to the third resonance region and for $Q^2
< 1\,(GeV/c)^2$.  In Fig. 1 we present our results for the total cross
sections at $Q^2=0$ and $0.5\,(GeV/c)^2$.

We have not yet included the contributions of the eta and multipion
channels to the longitudinal part of the cross section. Since the
ratio $R$ is small in the resonance region,
$R=\sigma_L/\sigma_T\approx 0.06$~\cite{Bra76,Abe97}, these
contributions should not be of major importance anyway.  On the other
hand, we find a substantial longitudinal strength for single-pion
production at the higher energies,
$R^{(1\pi)}=\sigma^{(1\pi)}_L/\sigma^{(1\pi)}_T>0.5$ at $W>1.8\, GeV$
and $Q^2>0.3\,(GeV/c)^2$. Hence we are led to the conclusion that the
(small) longitudinal contribution is dominated by single-pion
production.
        
\subsection{Asymmetries and structure functions}

One of the basic ingredients for the asymmetries $A_{1,2}$ and structure
functions $g_{1,2}$ is the difference of the two helicity  cross sections,  

\begin{eqnarray}
\label{eq16}
\Delta\sigma=\sigma_{3/2}-\sigma_{1/2}=\Delta\sigma^{(1\pi)}+
\Delta\sigma^{(\eta)} + \Delta\sigma^{(n\pi)}\,. 
\end{eqnarray} 

The single-pion contribution to this difference has been calculated
within the UIM approach. Since the dominant mechanism of eta
electroproduction is due to the excitation of the $S_{11}(1535)$
resonance, the contribution of the eta channel to Eq.~(\ref{eq16}) may
be approximated by $\Delta\sigma^{(\eta)} \approx
-2\sigma_T^{(\eta)}$. The description of the multipion channels is the
most delicate part in our calculation of the sum rules, because the
mentioned theoretical models~\cite{Gom94,Mur95,Och97} for two-pion
photoproduction were tuned to the existing data for total cross
sections.  The simplest estimate for $\Delta\sigma^{(n\pi)}$ is based
on the assumption that the two-pion contribution is generated by
resonances, and that its helicity structure follows the known behavior
of the one-pion contribution
$\Delta\sigma^{(1\pi)}$~\cite{Kar73,Lvo97}, i.e.
$\Delta\sigma^{(n\pi)}/\Delta\sigma^{(1\pi)}= const.$ In the present
paper we first account for the one-pion and eta channels explicitly,
and then use the simple prescription

\begin{eqnarray}
\label{eq17}
\Delta\sigma^{(n\pi)}= \frac{\sigma_T^{(n\pi)}}{\sigma_T}
\left(\Delta\sigma^{(1\pi)}+\Delta\sigma^{(\eta)}\right)\ ,
\end{eqnarray}  

with the factor $\sigma_T^{(n\pi)}$ in the numerator providing the
correct threshold behaviour for the two-pion contribution.

In the upper part of Fig. 2, our results are compared to the asymmetry
$A_1+\eta A_2$ measured at SLAC~\cite{Abe97}. As may be seen, the
prescription of Eq.~(\ref{eq17}) is in reasonable agreement with the
data up to $W=2\, GeV$.  We also note that the contribution of the
$\eta$ channel (dotted curves) leads to a substantial increase of the
asymmetry over a wide energy region.

Having fixed all the ingredients we can now calculate the structure
functions and generalized GDH integrals.  Our results for the
structure function $g_1$ are presented in the lower part of Fig. 2. Up
to a value of $W^2=2\, GeV^2$ (corresponding to $x=0.31$ and $x=0.52$
at $Q^2$=0.5 and 1.2 $(GeV/c)^2$, respectively), the main contribution
to $g_1$ is due to single-pion production.  The negative structure
above threshold is related to excitation of the $\Delta(1232)$
resonance. In the second and third resonance regions the contributions
from the eta and multipion channels become increasingly important.
Note that with increasing value of $Q^2$ the relative importance of
the eta channel increases. This phenomenon is connected with the well
known peculiarity that the helicity amplitude of the $S_{11}(1535)$
resonance decreases much slower with $Q^2$ than for any other
resonance~\cite{Bra84,Dre98a}.

\subsection{Integrals $I_1$ and $I_2$} 

In Fig. 3 we give our predictions for the integrals $I_1(Q^2)$ and
$I_2(Q^2)$ in the resonance region, i.e.  integrated up to
$W_{max}=2\,GeV$. The striking feature of the integral $I_1$ is the
evolution from large negative to small positive values in order to
interpolate from the GDH prediction for real photons to the data
obtained from DIS.  As can be seen from the top of Fig. 3, our model
is able to generate the dramatic change in the helicity structure
quite well.  While this effect is basically due to the single-pion
component predicted by the UIM, the eta and multipion channels are
quite essential to shift the zero-crossing of $I_1$ from
$Q^2=0.75\,(GeV/c)^2$ to 0.52 $(GeV/c)^2$ and 0.45 $(GeV/c)^2$,
respectively.  This improves the agreement with the SLAC
data~\cite{Abe98}.  However, some differences remain. Due to a lack of
data in the $\Delta$ region (see lower part of Fig. 2), the SLAC data
are likely to underestimate the $\Delta$ contribution. Since this
contribution is negative, a numerical integration of the data points
could overestimate the $I_1$ integral or the corresponding first
moment $\Gamma_1$. A few more data points in the $\Delta$ region would
be very useful in order to clarify the situation.

Concerning the integral $I_2$, our full results are in good agreement with
the predictions of the BC sum rule (see Fig. 3, lower part).  The
remaining differences are of the order of $10~\%$ and should be
attributed to contributions beyond $W_{\rm max}=2$ GeV and the scarce
experimental data for $\sigma'_{LT}$. With regard to the latter
problem we note that the BC sum rule, Eqs.~(\ref{eq7}) and
(\ref{eq10}), contains very sizeable contributions from the
longitudinal-transverse cross section.  In Table 1 we list the
contributions of the different ingredients of our model to the
integrals $I_1$ and $I_2$ at the real photon point.

The convergence of the sum rules can not be given for granted. In fact
Ioffe et al. \cite{Iof84} have argued that the BC sum rule is valid
only in the scaling region, while it is violated by higher twist terms 
at low $Q^2$. Therefore the good agreement of our model with the BC
sum rule could be accidental and due to a particular model prediction
for the essentially unknown longitudinal-transverse interference
term. As can be seen from Fig. 3, the contribution of $\sigma_{LT'}$
is quite substantial for $I_2$ even at the real photon point due to
the factor $\gamma^{-1}$ in Eq. 10.
This contribution, however, is constrained by the positivity relation 
$|\sigma_{LT}^{'}|\leq \sqrt{\sigma_L\sigma_T}=\sqrt{R}\,\sigma_T$. The
dash-dotted line shows the integral for the upper limit of this
inequality and a similar effect would occur for the lower limit. This
surprisingly large contribution can be understood in terms of multipoles.
In a realistic description of the integrated cross section
$\sigma_{LT'}$ the large $M_{1+}$ multipole can only
interfere with the small $L_{1+}$ multipole. The upper and lower
limits of the positivity relation overestimate the structure function
considerably due to an unphysical ``interference'' between $s$- and
$p$-waves. 

In  Fig. 4 we compare our  result for the function $\Gamma_1$, defined
in Eq.  (6),  with  the calculation  of  Ioffe  \cite{Iof97}  for  the
resonance contribution. This  representation demonstrates more clearly
the importance of the higher channels, e.g. $\eta$ and $2\pi$.

Fig. 5 shows our predictions for
$\Delta\sigma=\sigma_{3/2}-\sigma_{1/2}$ at $Q^2$=0, 0.5 and 1.2
$(GeV/c)^2$.  In particular we note the following peculiarity of the
$Q^2$ dependence of the resonance and multipion contributions.  With
increasing value of $Q^2$ the positive contribution of the $\Delta$
resonance decreases rapidly due to the strong $N\Delta$ transition
form factor.  In the second and third resonance regions, however,
$\Delta\sigma$ changes from positive to negative values of $Q^2$. The
reasons for this zero-crossing are (I) the helicity structure of the
$D_{13}$ and $F_{15}$ resonances, which changes from $\sigma_{3/2}$ to
$\sigma_{1/2}$ dominance~\cite{Bur88}, and (II) the surprisingly slow
fall-off of the $S_{11}$ transition form factor~\cite{Bra84,Dre98a},
which increases the negative contribution of this resonances relative
to the others.

Both effects conspire to cancel the $\Delta$ contribution already at
$Q^2\approx 0.5$(GeV/c)$^2$ and lead to the observed positive value of
$I_1$ at larger momentum transfer. The same cancellations are
responsible for the rapid decrease of $I_2$ with $Q^2$.  However, it
should be kept in mind that this integral also receives strong
contributions from $\sigma'_{LT}$. In view of our bad knowledge of the
longitudinal-transverse response we find it actually quite surprising
that the UIM ``knows'' about the BC sum rule and the predicted rapid
fall-off with $Q^2$.
      
\section{Conclusion}

We have studied the nucleon spin structure functions and the
generalized GDH and BC integrals for small and moderate momentum
transfer within the framework of the Unitary Isobar
Model~\cite{Dre98a}. The contributions from eta and multipion channels
in the second and third resonance regions have been modeled by
accounting for the total photoproduction cross section.
   
We find that the role of the eta channel becomes more and more
important with increasing values of $Q^2$. At the real photon point,
this channel tends to cancel the contribution of multipion production.
However, at finite $Q^2$ the contributions of both channels add
coherently. As a result the zero-crossing of the integral $I_1$ shifts
from $Q^2=0.75\,(GeV/c)^2$ (for the one-pion channel only) to
$Q^2=0.45\,(GeV/c)^2$ (if eta and multipion channels are included).

Our analysis indicates that both the measured asymmetries and the
theoretical models have to be quite accurate in order to determine the
integrals $I_1$ and $I_2$ in the resonance region.  As far as theory
is concerned, a more quantitative description of the multipion
channels is absolutely necessary. On the side of the experiment, a
measurement of all four virtual photon cross sections is needed to
determine the quark spin structure functions in a model-independent
way. Moreover, the oscillatory behaviour of the integrand in the
resonance region will require small error bars and a relatively small
energy binning in order to give precise values for the GDH and BC
integrals. The proposed experiments at Jefferson Lab~\cite{Bur91}
carry the promise to obtain such data in the resonance region and we
are eagerly waiting for the outcome of these investigations.

\begin{table}[htbp]
\begin{center}
\begin{tabular}{lllllll}
 $I_{1,2}$ & Born+$\Delta$  & $P_{11},D_{13},...$    
& $\eta$ & multipion & total & sum rule \\
\hline
 $I_1$ & -0.565 & -0.152 & 0.059  & -0.088 & -0.746   & -0.804\\
 $I_2$ & 1.246  & 0.063 &  -0.059  &   0.088 & 1.338 & 1.252\\
\end{tabular}
\end{center}  \caption{Contributions of the different channels
to the integrals $I_1$ and $I_2$  at the real photon point, $Q^2=0$.
The column "$P_{11},D_{13},...$" 
shows the contributions of all the resonances above the
$\Delta$ that are included in Ref.\protect\cite{Dre98a}. Note
that the column ``total'' lists the values for integration up to
$W_{\rm max}=2$ GeV only, while ``sum rule'' is the prediciton
for the full integral.
}
\end{table}

\begin{figure}[h]
\centerline{\psfig{file=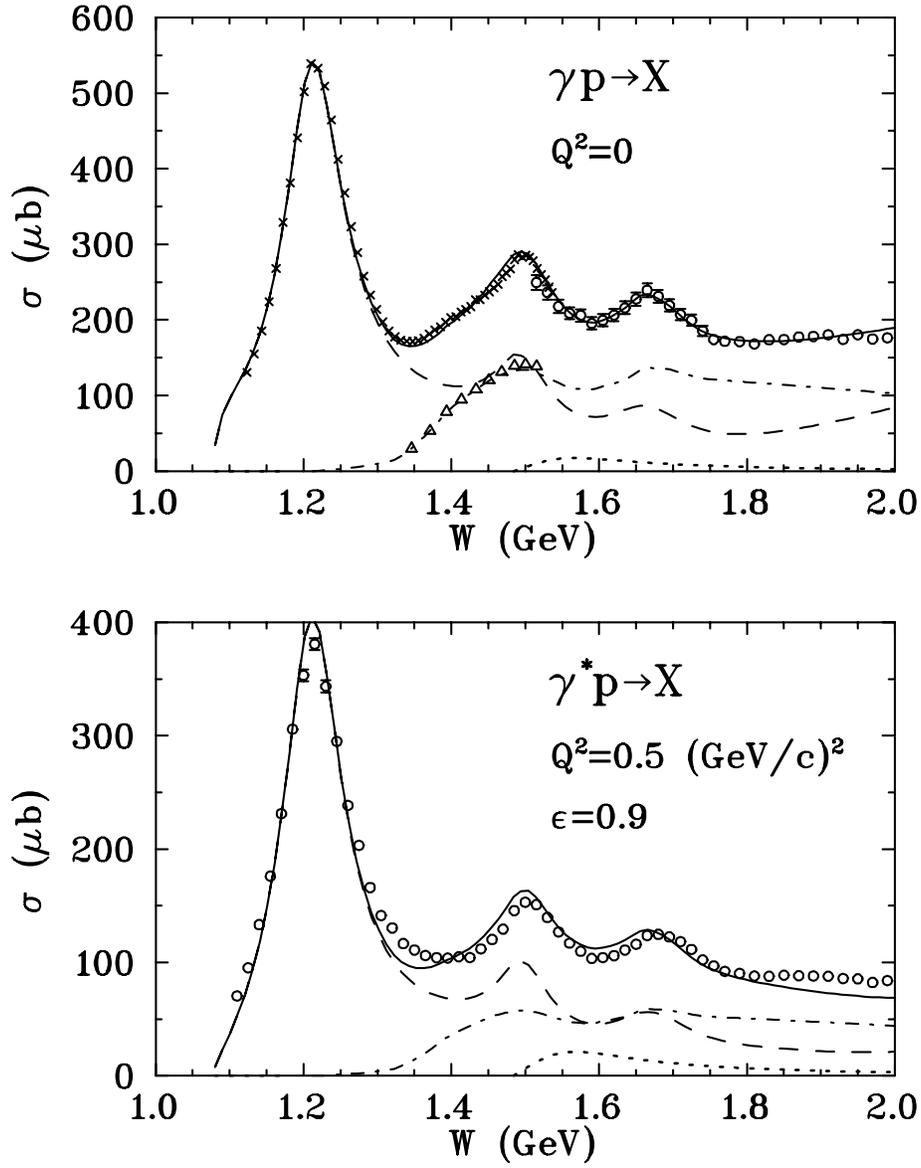,width=12cm,silent=}}
\vspace{0.5cm}
\caption{ Total cross sections for photoabsorption and inelastic
  electron scattering on the proton for $\varepsilon=0.9$ and
  $Q^2=0.5\,(GeV/c)^2$.  Dashed, dotted and dash-dotted curves:
  contributions of single-pion, eta and multipion channels,
  respectively; solid curves: final result. Experimental data for the
  total cross sections from Refs.\protect\cite{Bra76} (x) and
  \protect\cite{Mac96} ($\circ$), for the two-pion production channels
  from Ref.\protect\cite{Mac96} ($\bigtriangleup$).  }
\end{figure}

\begin{figure}[h]
\centerline{\psfig{file=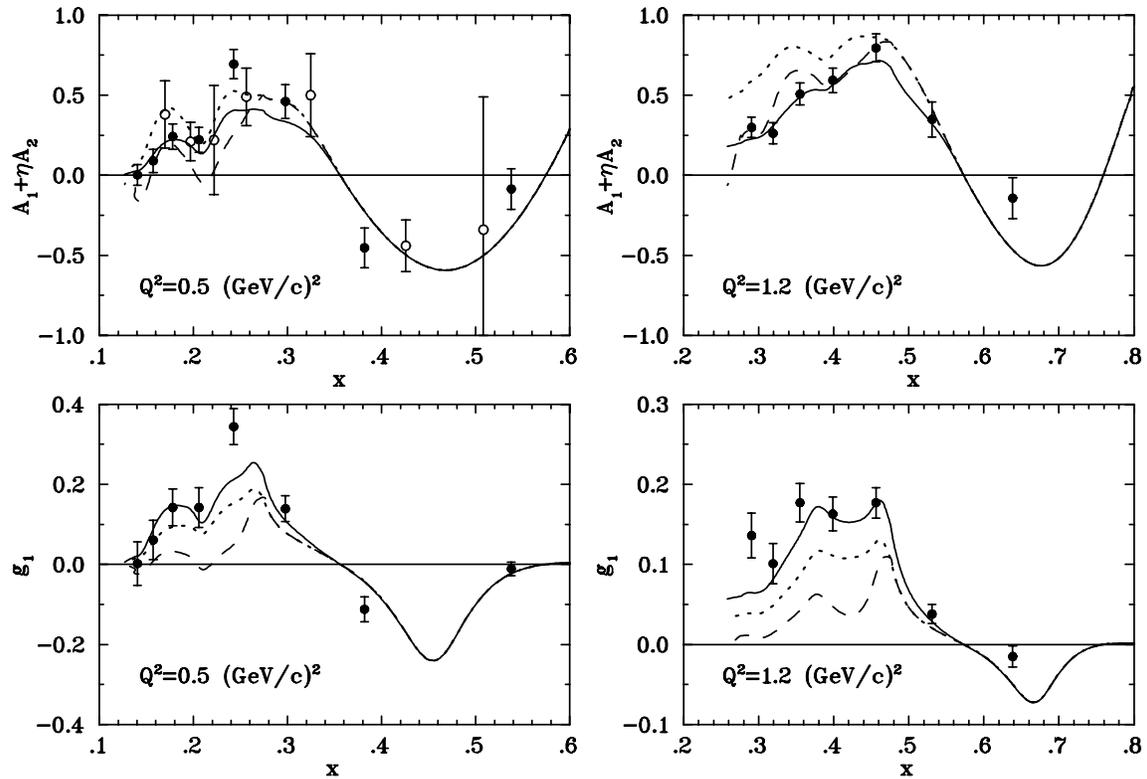,width=15cm,angle=90,silent=}}
\vspace{0.5cm}
\caption{
  The asymmetry $A_1+\eta A_2$ (top) and the spin structure function
  $g_1$ (bottom) as function of the Bjorken scaling variable $x$ at
  $Q^2=0.5$ and 1.2 $(GeV/c)^2$. Dashed, dotted and solid curves:
  calculations obtained with $1\pi$, $1\pi+\eta$, and
  $1\pi+\eta+n\pi$ contributions, respectively. Data from
  Refs.\protect\cite{Abe98} ($\bullet$) and \protect\cite{Bau80}
  $(\circ)$. The error bars give the statistical errors, the
  systematical errors are estimated to be of equal size.  }
\end{figure}

\begin{figure}[h]
\centerline{\psfig{file=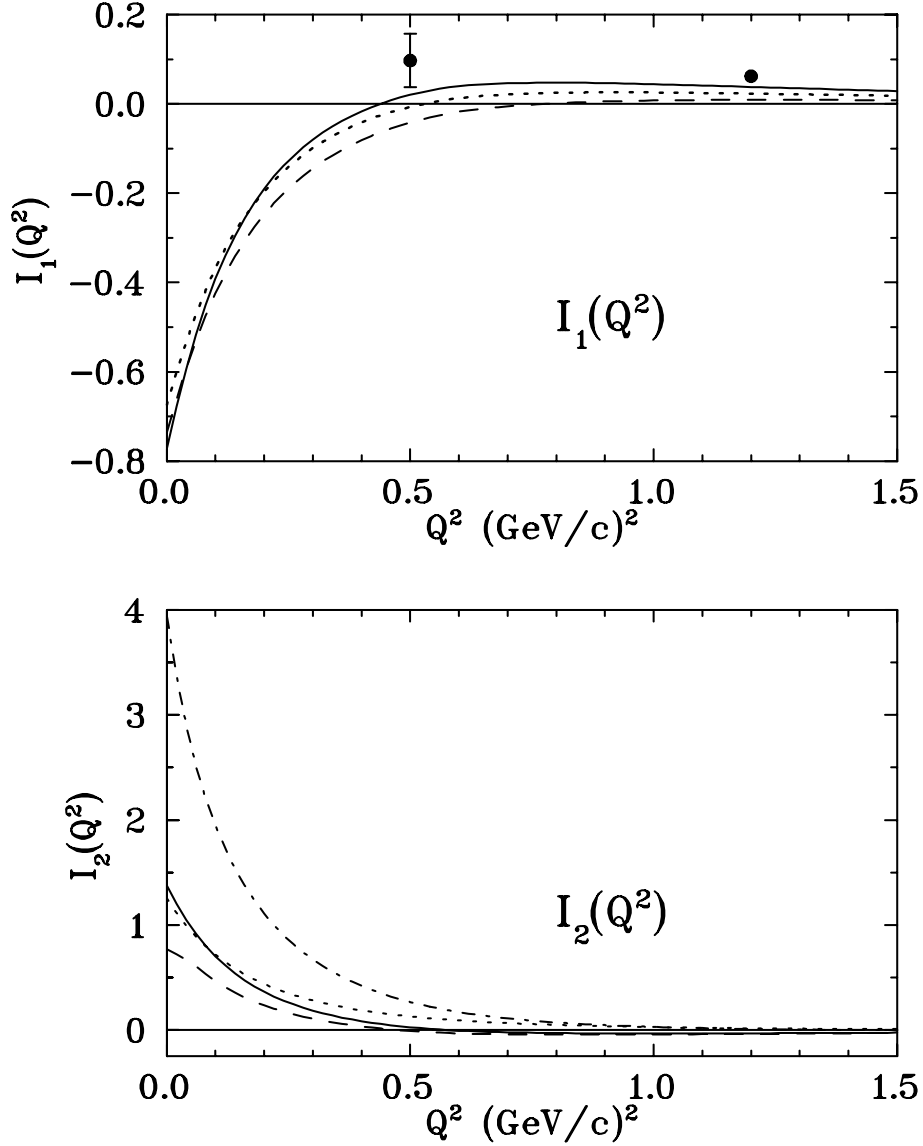,width=12cm,silent=}}
\vspace{0.5cm}
\caption{
  The integrals $I_1$ and $I_2$ defined by Eqs. (10) and (11) as
  functions of $Q^2$ in the resonance region, integrated up to
  $W_{max}=2\,GeV$.  Upper figure: Notation as in Fig. 2 and data from
  Ref.\protect\cite{Abe98}.  Lower figure: The full and dashed lines
  are our predictions with and without $\sigma_{LT}^{'}$ (see Eq. (10)),
  the dash-dotted line is obtained for
  $\sigma_{LT}^{'}=\sqrt{\sigma_L\sigma_T}$ (see text), and the dotted
  line is the sum rule prediction of Ref.\protect\cite{Bur70}. All
  calculations for $I_2$ include $1\pi + \eta + n\pi$ contributions.
  }
\end{figure}

\begin{figure}[h]
\centerline{\psfig{file=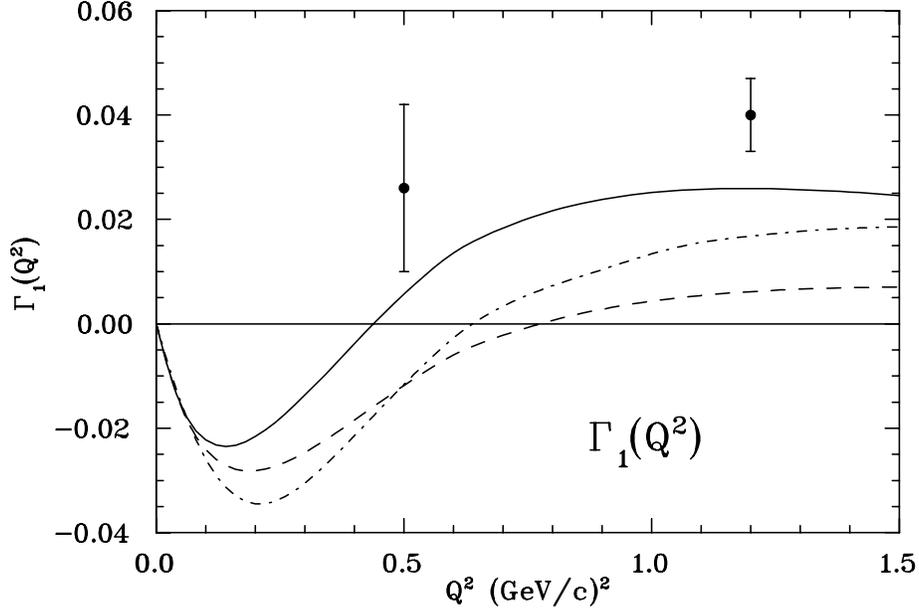,width=12cm,angle=90,silent=}}
\vspace{0.5cm}
\caption{
  The integral $\Gamma_1$ for the proton defined by Eq. (6) as
  function of $Q^2$. The solid and dashed lines show the integral up
  to $W_{max}=2 GeV$ for our full calculation and our calculation
  without the $\eta$ and $2\pi$ contributions, respectively. The
  dash-dotted line is obtained from Ref. \protect\cite{Iof97} and
  contains only the resonance component. The data are from
  Ref.\protect\cite{Abe98}.}
\end{figure}

\begin{figure}[h]
\centerline{\psfig{file=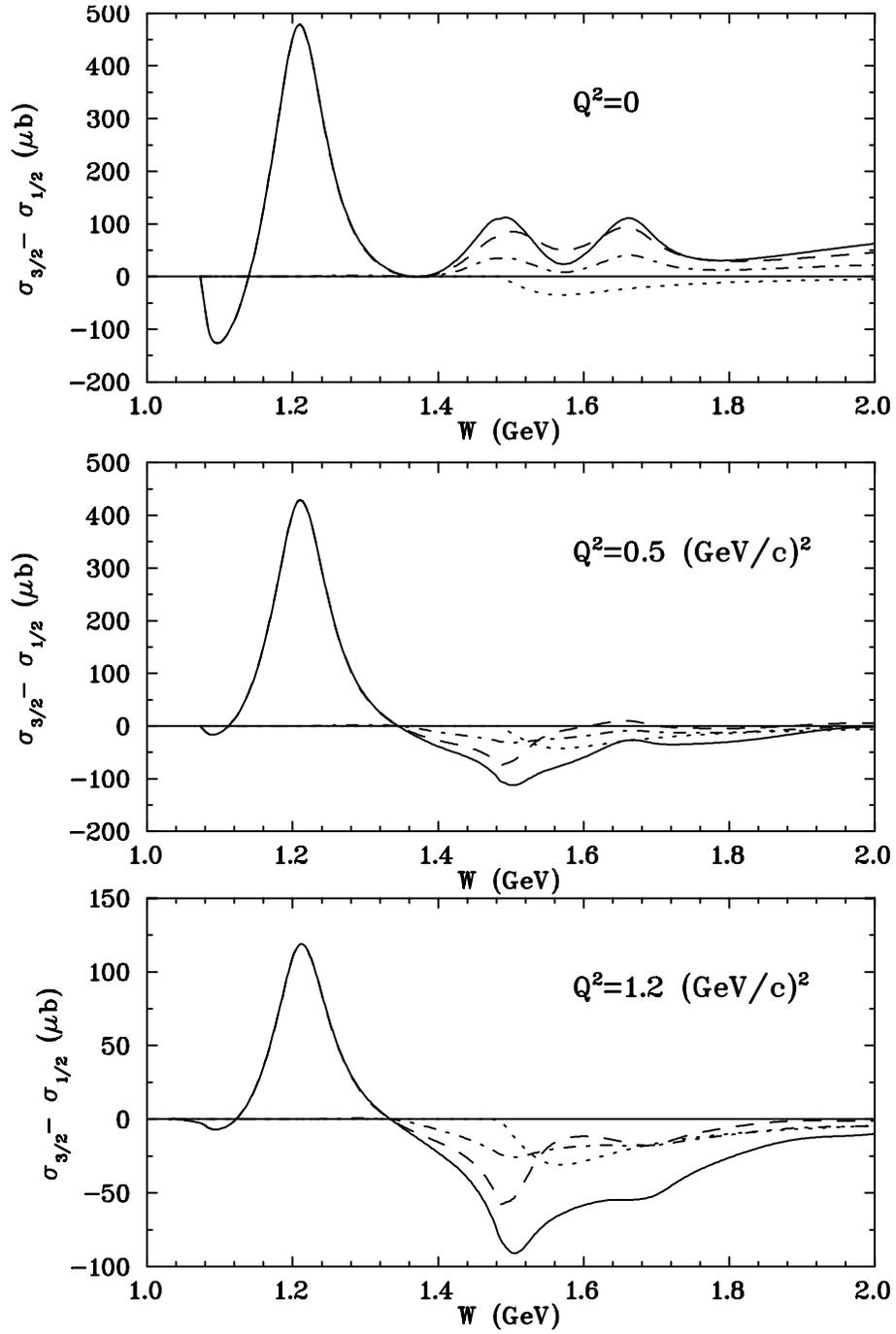,width=12cm,silent=}}
\vspace{0.5cm}
\caption{
  Predictions for $\Delta\sigma=\sigma_{3/2}-\sigma_{1/2}$ at $Q^2$=0,
  0.5 and 1.2 $(GeV/c)^2$.  Notation as in Fig. 1.  }
\end{figure}

\end{document}